\documentclass[12pt]{iopart}
\usepackage{graphicx}

\begin{document}

\title[Doping dependence of the $(\pi,\pi)$ shadow band in La214]{Doping dependence of the $(\pi,\pi)$ shadow band in La-based cuprates studied by angle-resolved photoemission spectroscopy}

\author{R.-H. He$^{1, 2, 3}$, X. J. Zhou$^{1, 3, 4}$, M. Hashimoto$^{1, 2, 3}$, T. Yoshida$^{1, 5}$, K. Tanaka$^{1, 2, 3, 6}$, S.-K. Mo$^{1, 2, 3}$, T. Sasagawa$^{1, 2, 7}$, N. Mannella$^{1, 2, 3, 8}$, W. Meevasana$^{1, 2, 9}$, H. Yao$^1$, M. Fujita$^{10}$, T. Adachi$^{11}$, S. Komiya$^{12}$, S. Uchida$^5$, Y. Ando$^{13}$, F. Zhou$^4$, Z. X. Zhao$^4$, A. Fujimori$^5$, Y. Koike$^{11}$, K. Yamada$^{10}$, Z. Hussain$^3$ and Z.-X. Shen$^{1, 2}$}

\address{$^1$Geballe Laboratory for Advanced Materials, Departments of Physics and Applied Physics, Stanford University, Stanford, California 94305, USA}
\address{$^2$Stanford Institute for Materials and Energy Sciences, SLAC National Accelerator Laboratory, Menlo Park, California 94025, USA}
\address{$^3$Advanced Light Source, Lawrence Berkeley National Laboratory, Berkeley, California 94720, USA}
\address{$^4$National Laboratory for Superconductivity, Beijing National Laboratory for Condensed Matter Physics, Institute of Physics, Chinese Academy of Sciences, Beijing 100190, China}
\address{$^5$Department of Physics, University of Tokyo, Bunkyo-ku, Tokyo 113-0033, Japan}
\address{$^6$Department of Physics, Osaka University, Osaka 560-0043, Japan}
\address{$^7$Materials and Structures Laboratory, Tokyo Institute of Technology, Kanagawa 226-8503, Japan}
\address{$^8$Department of Physics and Astronomy, University of Tennessee, Knoxville, Tennessee 37996, USA}
\address{$^9$School of Physics, Suranaree University of Technology, Nakhon Ratchasima, 30000 Thailand}
\address{$^{10}$Institute of Materials Research and World-Premier-International Research Center Initiative, Tohoku
University, Sendai 980-8577, Japan}
\address{$^{11}$Department of Physics and Applied Physics, Tohoku University, Sendai 980-8579, Japan}
\address{$^{12}$Central Research Institute of Electric Power Industry, Komae, Tokyo 201-8511, Japan}
\address{$^{13}$Institute of Scientific and Industrial Research, Osaka University, Ibaraki, Osaka 567-0047, Japan}

\begin{abstract}
The $(\pi,\pi)$ shadow band (SB) in La-based cuprate family (La214) was studied by angle-resolved photoemission spectroscopy (ARPES) over a wide doping range from $x=0.01$ to $x=0.25$. Unlike the well-studied case of the Bi-based cuprate family, an overall strong, monotonic doping dependence of the SB intensity at the Fermi level ($E_F$) was observed. In contrast to a previous report for the presence of the SB only close to $x=1/8$, we found it exists in a wide doping range, associated with a doping-independent $(\pi,\pi)$ wave vector but strongly doping-dependent intensity: It is the strongest at $x\sim 0.03$ and systematically diminishes as the doping increases until it becomes negligible in the overdoped regime. This SB with the observed doping dependence of intensity can in principle be caused by the antiferromagnetic fluctuations or a particular form of low-temperature orthorhombic lattice distortion known to persist up to $x\sim 0.21$ in the system, with both being weakened with increasing doping. However, a detailed binding energy dependent analysis of the SB at $x=0.07$ does not appear to support the former interpretation, leaving the latter as a more plausible candidate, despite a challenge in quantitatively linking the doping dependences of the SB intensity and the magnitude of the lattice distortion. Our finding highlights the necessity of a careful and global consideration of the inherent structural complications for correctly understanding the cuprate Fermiology and its microscopic implication. 
\end{abstract}

\pacs{74.25.Jb, 74.72.Dn, 79.60.-i}
\maketitle

\section{Introduction}

A current focus of high-$T_c$ superconductivity is on the topology of the Fermi surface of underdoped cuprates in the normal state well above $T_c$ (see Ref. \cite{HTSC:InnaReview} and other papers in the same focus issue). Whether it is a single, large, hole-like Fermi surface or consists of multiple, small, electron- and hole-like Fermi surfaces, determines whether high-$T_c$ superconductivity is an instability of a gapless metallic state with a single-band band structure \cite{cuprates:Bi2212:NodalMetal2} or its partially-gapped variant with a transformed band structure due to some lattice (rotational and/or translational) symmetry breaking \cite{HTSC:PseudogapDispersion}. Central to this debate is the existence and origins of SBs, which represent additional ingredients in the band structure that goes beyond the single-band picture.

The SB is defined in relation to the main band (MB). It generally has weaker spectral weight and holds a certain symmetry relation with the MB. Its appearance in ARPES \cite{misc:PES:ZXreview} indicates the existence of some form of super-modulation (of wave vector $\vec{q}$) in the crystal that scatters the electronic states from the MB at $\vec{k}$ to the SB at $\vec{k}+\vec{q}$. Depending on the form of super-modulation, it usually has a structural \cite{cuprates:Bi2212:Sbpipi_3}, electronic \cite{CDW:VeronqiueReview} or magnetic \cite{CDW:SDW_Cr_Eli} origin, as have been demonstrated in various materials. Concerning high-$T_c$ superconductivity, the electronic or magnetic super-modulation, as the result of collective excitations of the many-body electronic system, has been generally deemed to be more relevant. 

Particularly, about the SB with $\vec{q}=(\pi,\pi)$ in Bi-based cuprates, consensus was finally reached after years of debate between scenarios for its magnetic \cite{cuprates:Bi2212:Sbpipi_5} or structural \cite{cuprates:Bi2212:Sbpipi_6, cuprates:Bi2212:Sbpipi_7} origin. The lack of additional renormalization in the dispersion and lack of additional momentum broadening of the SB relative to the MB, the binding energy independence of the intensity ratio of the SB over the MB (SB/MB) \cite{cuprates:Bi2212:Sbpipi_2}, all argue against the magnetic interpretation where the antiferromagnetic correlations in this hole-doped system are generally recognized to be dynamic rather than static. This intensity ratio was found to exhibit a doping dependence in the underdoped regime that is opposite \cite{cuprates:Bi2212:Sbpipi_1} (if finite \cite{cuprates:Bi2212:Sbpipi_1_2}) to the expectation for the weakening of antiferromagnetic correlations as doping increases. On the other hand, the structural alternative was confirmed by a polarization dependence study of untwinned samples, which uncovered the hidden four-fold symmetry breaking nature of the SB formation \cite{cuprates:Bi2212:Sbpipi_4}. This SB has thus been concluded to arise from a unique form of (bulk) orthorhombic distortion of a tetragonal lattice structure that breaks its original four-fold symmetry and primarily takes place in the BiO planes.

Such a structural interpretation unique to Bi-based cuprates apparently cannot be directly applied to explain a similar existence of $(\pi,\pi)$ SBs at various doping levels of La214 reported in literatures \cite{cuprates:LSCO:XingjiangDichotomy, cuprates:LSCO:Teppei_LSCOFS, cuprates:Bi2201:Sbpipi_1, cuprates:LSCO:TeppeiReview, cuprates:LSCO:XingjiangBookChap, cuprates:LSCO:SBpipi_Mesot}. However, in the low-temperature orthorhombic (LTO) phase of La214 \cite{stripe:neutron:LSCOSpinIncomm_vertical, cuprates:LSCO:SBpipi_LTOHTT1} and also locally in the high-temperature tetragonal (HTT) phase above \cite{cuprates:LSCO:SBpipi_LTOHTT2}, another form of $(\pi,\pi)$ orthorhombic lattice distortion exists, which is caused by a staggered tilting of the CuO$_6$ octahedra around the Cu-O bond diagonal. While the contribution of this structural aspect might be similarly expected to dominate the SB formation in La214, a recent report \cite{cuprates:LSCO:SBpipi_Mesot} of an anomalous enhancement of the SB intensity at $x=1/8$ argued for its intimate connection with the coincident static spin and charge stripe orders, despite the fact that these orders are associated with some incommensurate wave vectors different from $(\pi,\pi)$ \cite{stripe:neutron:LBCO_Tranquada, stripe:neutron:LBCO_Fujita, stripe:other:LBCO_Xray}. Such anomaly tied to $x=1/8$, if it turns out to be true, would strongly support the multi-band nature of the band structure of La214 in the normal state and its origin due to a specific form of lattice symmetry breaking. However,  a considerable intensity of this SB has been reported at other doping levels \cite{cuprates:LSCO:XingjiangDichotomy, cuprates:LSCO:Teppei_LSCOFS, cuprates:Bi2201:Sbpipi_1, cuprates:LSCO:TeppeiReview, cuprates:LSCO:XingjiangBookChap}, which casts such anomaly into doubt. Before a reliable evaluation of its potential importance for a unified understanding of the origin of the SB in cuprates can be made, a careful re-examination of this possible anomaly in La214 has to be performed, which should be placed in a wider context by means of a more systematic doping dependence study in La214.

In this paper, we present ARPES data at doping levels ranging from $x=0.01$ to $x=0.25$ in various members of La214. Unlike the Bi-based cuprate family, a strong, monotonic doping dependence of the SB intensity at the Fermi level ($E_F$) was observed. Although the SB intensity fluctuates by a considerable amount from experiments to experiments at a given nominal doping level, we found it is generally stronger at lower dopings and systematically weakens with the increase of doping until it becomes negligible in the overdoped regime. The observed existence of the SB over a wide doping range and its overall weakening of intensity are very different from the previous report for a maximal SB intensity at $x=1/8$ and negligible intensity at other dopings \cite{cuprates:LSCO:SBpipi_Mesot}. While such trend of the SB intensity appears to be consistent with a decreasing strength of antiferromagnetic fluctuations in the system, similar lacks of additional renormalization in the SB dispersion and of binding energy dependence of the SB/MB intensity ratio are found and exemplified at $x=0.07$, which questions the conventional magnetic interpretation as proposed for the Bi-based case \cite{cuprates:Bi2212:Sbpipi_2, cuprates:Bi2212:Sbpipi_8}. On the other hand, the observation that the SB becomes virtually undetectable near $x\sim 0.21$ is broadly consistent with the structural interpretation based on the aforementioned orthorhombic distortion which disappears at $x\sim 0.21$. However, important discrepancies in their detailed doping dependences are noted, which pose a challenge for establishing their precise connection in a quantitative fashion.

\section{Experimental}

ARPES measurements were carried out at Beam Line 10.0.1 in Advanced Light Source (ALS) using different versions of Scienta electron energy analyzer, SES-200, SES-2002 and SES-R4000, over the years (1999 - present). The photon
energies ($h\nu$) used are 55, 55.5 \& 59.5 eV and the polarization of light was fixed in plane and orthogonal to the zone diagonal (nodal) direction during each experiment by virtue of a sample-fixed analyzer-rotating capability. The typical energy resolution was $\sim$20 meV and angular resolution was $\sim$ 0.25$^\circ$ (0.015\AA$^{-1}$ in momentum). Single crystals grown by the travelling solvent floating zone method were studied for different members of La214 family, La$_{2-x}$Sr$_x$CuO$_4$ (LSCO or LS, grown by groups in Tokyo Institute of Technology \cite{cuprates:LSCO:TakaoLSCOGrowth_1, cuprates:LSCO:TakaoLSCOGrowth_2}, University of Tokyo \cite{cuprates:LSCO:UchidaLSCOGrowth}, Electric Power Industry \cite{cuprates:LSCO:AndoLSCOGrowth}, Chinese Academy of Sciences \cite{cuprates:LSCO:ZXZhaoLSCOGrowth}), La$_{2-x}$Ba$_x$CuO$_4$ (LBCO or LB, grown by groups in Tohoku University \cite{stripe:neutron:LBCO_Fujita, stripe:other:LBCO_Adachi, stripe:other:LBCO_Adachi_2}) and 1\% Fe-doped LSCO (Fe-LSCO or FeLS, grown by the group in Tohoku University \cite{stripe:neutron:FeLSCO2}). Superconducting transition temperature (T$_c$) vs. doping relations are similar to the published results (of LSCO \cite{stripe:neutron:LSCOSpinIncomm_vertical}, LBCO \cite{stripe:other:LBCO_Adachi_2}, Fe-LSCO \cite{stripe:other:WeaktoStrongCrossover}) but varies sightly among those grown by different groups. Samples were cleaved in vacuum with a typical base pressure better than 5x10$^{-11}$ Torr and measured at temperatures $\sim$20 K.

\begin{figure}
\begin{center}
\includegraphics[angle=0, width=3.2 in]{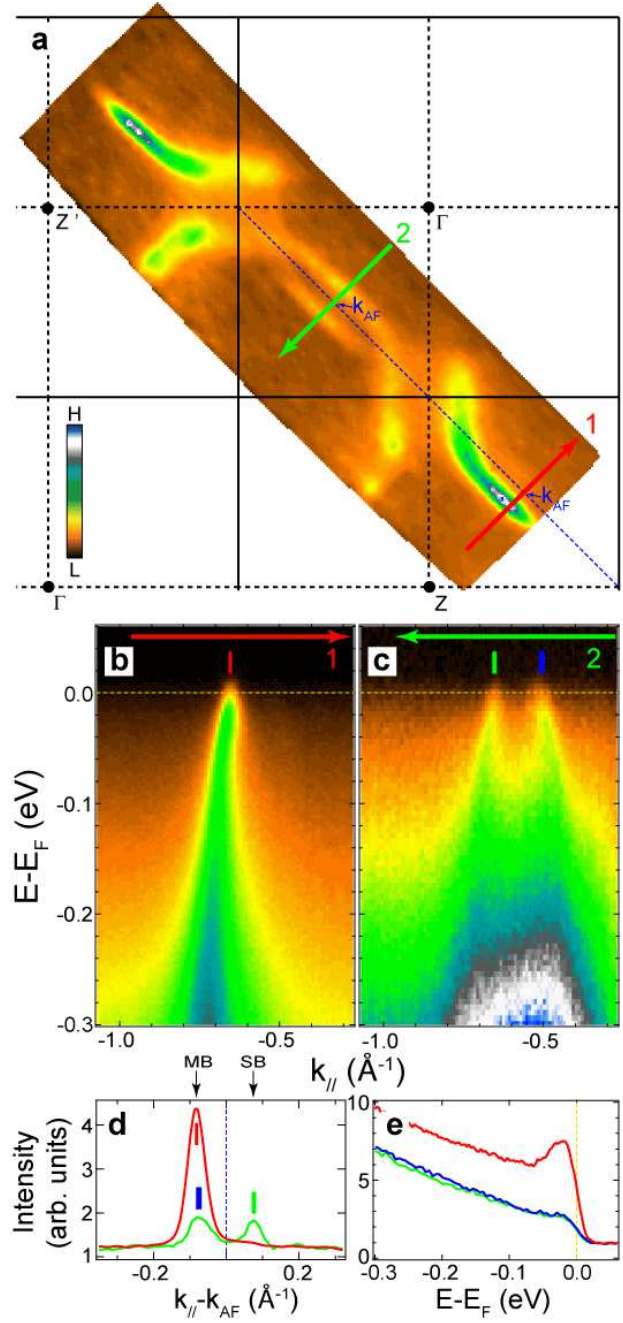}
\end{center}
\caption{\textbf{a}, FS map of LSCO $x=0.07$ measured by SES-2002 across 2 BZs (Z'-$\Gamma$-Z) taken with cuts parallel to the bond diagonal (nodal) direction. $h\nu$=55 eV. Intensity integration window is $E_F\pm$5 meV. Blue dashed line is the $(\pi,\pi)$ BZ boundary, whose crossing with a given cut is denoted as $k_{AF}$ hereafter. \textbf{b} \& \textbf{c}, ARPES intensity as a function of binding energy and parallel momentum along cuts 1 \& 2, the nodal direction in 2nd and 1st BZ, respectively. The momentum direction and range as indicated by the red and green arrows correspond to those in \textbf{a}. \textbf{d}, Intensity line profiles of cuts 1 \& 2 in \textbf{a} shown in corresponding colors. Red, blue and green vertical bars point out the Fermi crossings ($k_F$) of MB in 2nd BZ, MB and SB in 1st BZ, respectively, which correspond to those in \textbf{b} \& \textbf{c}. The bar thickness reflects the uncertainty in determining $k_F$. \textbf{e}, Energy distribution curves (EDCs) at $k_F$ marked by the same colors in \textbf{b} \& \textbf{c}.}
\label{Fig. 1}
\end{figure}

\section{Results and Discussion}
\subsection{\label{SB0d07} Existence of the $(\pi,\pi)$ shadow band at $x=0.07$}

Fig. \ref{Fig. 1}a shows an example for the typical Fermi surface (FS) mapping geometry of this study. The spectral intensity in the nodal region of 2nd Brillouin zone (BZ) is relatively enhanced compared to the one in 1st BZ due to the matrix element effects \cite{cuprates:LSCO:XingjiangDichotomy, cuprates:LSCO:TeppeiDichotomy}. Despite the suppression of overall intensity and quasi-particle spectral weight (Fig. \ref{Fig. 1}e), the spectra along the nodal cut in 1st BZ of LSCO $x=0.07$ show two counter-dispersing bands which have comparable spectral weight (Fig. \ref{Fig. 1}c). Their momemta are approximately symmetric about the $(\pi,\pi)$ BZ boundary (Fig. \ref{Fig. 1}d). We denote hereafter the ones marked in blue and in green as the MB and its $(\pi,\pi)$ SB, respectively. In contrast, the spectra taken at the nodal cut in 2nd BZ is dominated by a single dispersing band with barely noticeable intensity on its corresponding SB feature (Fig. \ref{Fig. 1}b \& d). The inequivalence of the SB/MB intensity ratio at equivalent in-plane momenta indicates the SB and MB have different matrix element effects. This can be more easily understood if the SB reflects an inherent symmetry of initial states (i.e., the SB exists in the band structure), similar as the Bi-based case \cite{cuprates:Bi2212:Sbpipi_4, Note_detwinned}, rather than being a trivial replica of the MB due to the final-state photoelectron diffraction.

\subsection{\label{MDCanalysis} MDC analysis of the $(\pi,\pi)$ shadow band at $x=0.07$}

\begin{figure}
\begin{center}
\includegraphics[angle=0, width=6 in]{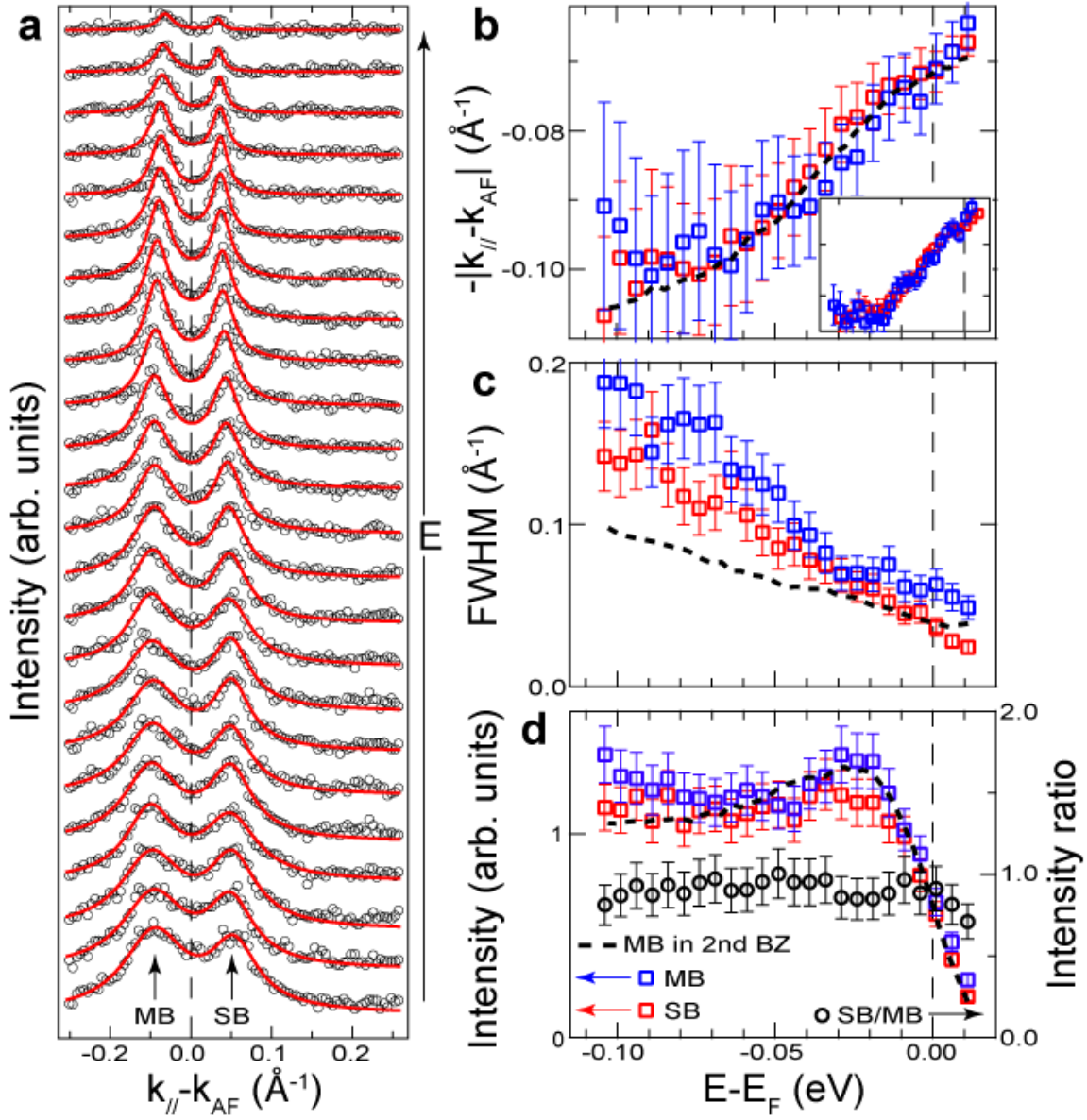}
\end{center}
\caption{\textbf{a}, MDCs from E-$E_F$=-104 meV (bottom) to 11 meV (top) shown in 5 meV step for Fig. \ref{Fig. 1}c. Red curves are MDC fits with two Lorentzians and a linear background. \textbf{b}-\textbf{d}, Results of the MDC analysis on the MDC peak dispersion, MDC peak width (FWHM) and MDC peak intensity, respectively, of the MB and SB in 1st BZ (\textbf{a}), in comparison with the results for the MB in 2nd BZ (Fig. \ref{Fig. 1}b; error bars not shown are slightly smaller than the former). In \textbf{b}, the SB dispersion is folded symmetrically about $k_{AF}$ for comparison. Error bars are from the MDC fitting. Inset: Results of fitting each peak individually with a Lorentzian and a linear background for the MB and SB in 1st BZ. The SB/MB intensity ratio is also shown in \textbf{d}.}
\label{Fig. 2}
\end{figure}

To check if a magnetic interpretation is compatible with the observed $(\pi,\pi)$ SB, we performed a momentum-distribution-curve (MDC) analysis similar as Ref. \cite{cuprates:Bi2212:Sbpipi_2} in Fig. \ref{Fig. 2}. At $x=0.07$, commensurate antiferromagnetic fluctuations have a broad maximum centered at the energy transfer position $\sim 25$meV according to Ref. \cite{stripe:neutron:LSCOHourGlassSaddleScaling}. As discussed in Ref. \cite{cuprates:Bi2212:Sbpipi_2} for the Bi-based case, if they are at play in our case, one expect the following observations of the SB (in comparison to the MB): i) an energy renormalization in the dispersion due to the fermionic exchange of spin fluctuations at finite energy transfers; ii) an additional momentum broadening of the order of 0.1 \AA$^{-1}$ due to the short-ranged nature of the fluctuations; iii) a binding energy dependence of the SB/MB intensity ratio by the same argument as i). Within experimental error bars, neither of the above predictions is supported by the results shown in Fig. \ref{Fig. 2}b-d, respectively, making the magnetic interpretation less likely a candidate for the cause of the SB.

Note that the existence of a kink-like feature in the SB dispersion at $\sim -25$meV is an artifact due to the imperfect fitting quality of the two-peak fit which can be improved by fitting the SB or MB peak feature individually (the inset of Fig. \ref{Fig. 2}b); the MDC peak width of the SB appears to be systematically smaller than that of the MB, which is beyond the error bars, robust against different MDC fitting schemes and quite commonly seen for other dopings (Fig. \ref{Fig. 3}a). Its trivial cause by the sample misalignment can be ruled out as many cuts, for example, Cut 2 (Fig. \ref{Fig. 1}a), are precisely aligned with the high-symmetry direction. 

Another puzzling observation was found when we performed the MDC analysis of the MB in 2nd BZ based on a single Lorentzian fit. While its peak dispersion and intensity are similar to those of the bands in 1st BZ within error bars (Fig. \ref{Fig. 2}b \& d), the peak width has a different energy dependence and the deviation is more pronounced towards higher binding energy (Fig. \ref{Fig. 2}c). This observation that both bands in 1st BZ tend to broaden more radically than that in 2nd BZ as a function of binding energy is also visible by eyes in Fig. \ref{Fig. 1}b \& c. We note the following. First, a similar observation has also been made at all other doping levels on which we could reliably perform a similar MDC analysis. Second, when one compares states in different 2D projected BZs, there are additional complications due to the effects of finite $k_z$ dispersion and matrix element. At this stage, we can fairly speculate at least two corresponding possibilities. There could exist a $k_z$-dependent scattering mechanism that causes a different broadening of states at different $k_z$. More interestingly, some additional nodal states located away from $E_F$ could be responsible for the additional energy broadening observed in 1st BZ but have an unfavorable photoemission matrix element in 2nd BZ with our experiment settings.

Both above anomalies in the MDC peak width of the SB and MB and of the bands in 1st and 2nd BZs, have not been previously reported, to our knowledge. Their origins are currently unclear, and probably deserve special attention and future studies.

\subsection{Doping dependence of the $(\pi,\pi)$ shadow band}

We have compiled data of comparable quality which were similarly taken at different doping levels of various members of La214, as shown in Figs. \ref{Fig. B200}-\ref{Fig. BR4k}. The $(\pi,\pi)$ SB can be seen for many doping levels, consistent with previous reports \cite{cuprates:LSCO:XingjiangDichotomy, cuprates:LSCO:Teppei_LSCOFS, cuprates:Bi2201:Sbpipi_1, cuprates:LSCO:TeppeiReview, cuprates:LSCO:XingjiangBookChap, cuprates:LSCO:SBpipi_Mesot}. If we simply focus on 1st BZ where its existence is the most obvious in the field of view for each FS map, we can see that it is generally more pronounced for low dopings than for high dopings, hinting at its finite doping dependence.

The simultaneous detection of the SB and MB within a single cut measured with identical instrumental settings (polarization, mapping geometry and $h\nu$), e.g., Cut 2 in Fig. \ref{Fig. 1}a, offers us a good chance to quantitatively assess the doping dependence of the SB intensity. The SB and MB can have different matrix elements but they are largely doping independent, as doping should not change the characters of states but mainly their weights and relative proportions. We thus normalize the intensity at $E_F$ of the SB by the MB in the same cut and directly compare this SB/MB intensity ratio, which mainly reflect the intrinsic spectral weight of the SB, between different experiments at different dopings. Such normalization scheme also allows us to bypass the difficulty in the spectral intensity normalization between different momentum cuts, e.g., cuts 1 \& 2 in Fig. \ref{Fig. 1}a, and to avoid most time-dependent instrumental artifacts due to the beam or/and manipulator instability, etc..

\begin{figure}
\begin{center}
\includegraphics[angle=0, width=2.9 in]{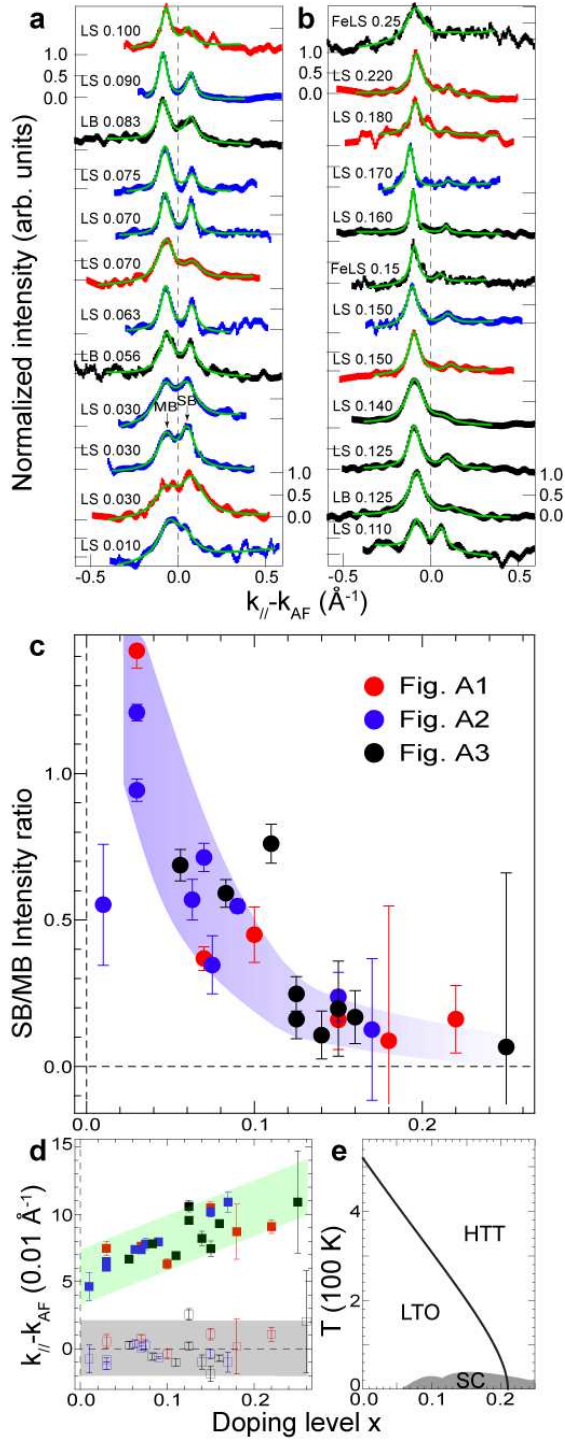}
\end{center}
\caption{\textbf{a} \& \textbf{b}, Maximum-normalized intensity line profiles along the nodal cut in 1st BZ (shown in green in Fig. \ref{Fig. B200}a) for all FS maps shown in Figs. \ref{Fig. B200}-\ref{Fig. BR4k}. Green curves are fits with two Lorentzians of identical FWHM \cite{Note_fit} and a linear background. \textbf{c}, The SB/MB intensity ratio at $E_F$ by the fitting. \textbf{d}, $k_F$ separation (measured from $k_{AF}$, solid symbol) and midpoint (empty symbol) between the SB and MB by the fitting. Error bars are from the fitting. Color ribbons are guides to the eye. \textbf{e}, Doping dependence of the HTT-LTO phase transition temperature and T$_c$ for LSCO, reproduced from Ref. \cite{cuprates:LSCO:SBpipi_LTOHTT2}.}
\label{Fig. 3}
\end{figure}

By using a similar two-peak MDC fitting routine, we obtain the doping dependence of this ratio (Fig. \ref{Fig. 3}c), the $k_F$ separation and midpoint between the SB and MB (Fig. \ref{Fig. 3}d). Within error bars, the midpoint is maintained close to $k_{AF}$, suggesting the wave vector associated with the SB is close to $(\pi,\pi)$ and independent of doping; the $k_F$ separation shows a systematic increase with doping, which is consistent with the increase of Luttinger's volume of the Fermi surface and the corresponding shift of the node position away from $k_{AF}$ \cite{cuprates:LSCO:Teppei_LSCOFS, stripe:other:WeaktoStrongCrossover}.

We will focus on the SB/MB intensity ratio at $E_F$ in the following. A salient feature of Fig. \ref{Fig. 3}c is that there exists quite large scattering in this ratio beyond the fitting error bars between different experiments on different samples at a given doping level. Because most instrumental artifacts aforementioned are expected to be eliminated by taking the intensity ratio, this result more likely reflects the uncertainties related to the samples, e.g., the variations in the surface doping, impurity concentration and/or cleaving quality, etc.. It in turn highlights the necessity of a systematic doping dependence study, with particular attention paid to the reproducibility of results, in order to obtain true insights about this problem. In this regard, the abrupt jumps observed at $x=0.01$ and $x=0.11$, each from only one sample, should be treated cautiously and be subject to more studies at these two doping levels, before a reliable conclusion on their intrinsic or extrinsic nature is reached.

Despite the pronounced scattering of results at a given doping level and the uncertain anomalies at $x=0.01$ and $x=0.11$, the intensity ratio shows an overall monotonic doping dependence. At low dopings, the SB intensity appears very strong, even comparable with the MB intensity at $x=0.03$, and has a strong doping dependence. Both the intensity and its doping dependence weaken with the increase of doping. At high dopings, the doping dependence, if finite, seems very weak. Affected by a noise level which is comparable to the weak SB intensity, the two-peak fitting routine tends to give finite peak intensity value in this regime. Although it is in principle uncertain with such large error bars (and scattering of results) whether the slope is zero or finite, we note that the SB can show up on some samples and at particular doping levels in this regime, e.g., at $x\sim 0.15$ in Fig. \ref{Fig. 3}b, which is consistent with what was reported in Ref. \cite{cuprates:Bi2201:Sbpipi_1}). 

Besides, from our data, it is difficult to tell a significant difference in the SB intensity between $x=1/8$ and above. In contrast, it was reported previously the SB intensity is enhanced at $x=1/8$, compared with both its lower and higher doping side where virtually no SB could be identified \cite{cuprates:LSCO:SBpipi_Mesot}. At $x<1/8$, the discrepancy with our data is even larger: Here we can clearly observe a systematically stronger SB intensity than at $x\ge 1/8$. We note that the previous result was based on a study of three doping levels, each with only one sample. In light of our preceding discussion, one should generally be cautious not to over-interpret results obtained in such a way. 

The strong doping dependence in the low-doping regime contrasts the Bi-based case whose doping dependence is subtle and still controversial (doping dependent \cite{cuprates:Bi2212:Sbpipi_1} or independent \cite{cuprates:Bi2212:Sbpipi_1_2}). Together with the weak doping dependence in the high-doping regime, this unique behavior presumably holds the clue to understand the origin of the $(\pi,\pi)$ SB in La214. While the magnetic scenario can likely be ruled out based on the binding energy dependence results at $x=0.07$, we are left with the structural alternative to be discussed below. 

The $(\pi,\pi)$ orthorhombic distortion sets in at the HTT-LTO phase transition which show an increased distortion magnitude as temperature decreases \cite{cuprates:LSCO:SBpipi_LTOHTT1}. As a result, the distortion magnitude at a fix low temperature, which determines how much spectral weight is transfered from the MB onto the SB, decreases with doping. This scenario generally expects a stronger SB intensity for a larger distortion magnitude and can thus explain the observed monotonic decrease of the intensity ratio in a qualitative fashion. It also anticipates no SB to be observable above $x\sim 0.21$, which is also consistent with the experiment. However, significantly strong distortions with a strong doping dependence have been found to persist up close to $x\sim 0.20$. For example, the phase line runs through the optimal doping almost linearly at $\sim$ 200 K, as shown in Fig. \ref{Fig. 3}e. This suggests a strong orthorhombic distortion at our measurement temperature and its strong doping variation, which are seemingly at odds with the weak SB intensity and its weak doping dependence observed at $x\sim 0.15$ and above. Nevertheless, we should remark that such a semi-quantitative assessment of the structural perspective for our experimental result might be premature. The exact relationship between the distortion magnitude and the intrinsic spectral weight of the SB is yet to be established, which might or might not be quadratic (as expected for a simple final-state photoelectron diffraction). Our observation previously discussed that the SB more likely arises from the scattering of the initial states, is relevant in this regard.

Although our finding suggests that the $(\pi,\pi)$ SB might not be directly connected with the stripe order as previously suggested \cite{cuprates:LSCO:SBpipi_Mesot}, it does not preclude the connection of the stripe order with the mysterious normal-state properties of cuprates. In fact, it has been recently shown that the stripe fluctuations, the precursor to the stripe order, set in below the pseudogap temperature $T^*$ for both the Y-based and La214 families \cite{HTSC:PseudogapBreakRotational}. A latest scanning tunnelling spectroscopy study supports a similar conclusion in Bi-based cuprates \cite{HTSC:PseudogapFluctuatingStripes_Yazdani}. Combined with our recent observation of the violation of particle-hole symmetry (as a hallmark of superconductivity) of the pseudogap \cite{HTSC:PseudogapDispersion}, it is therefore very likely that the normal-state band structure of cuprates is far more complicated than the single-band starting point that has been generally assumed before 2006 \cite{cuprates:Bi2212:Kiyo_TwoGap}. The band structure reconstruction due to the fluctuating stripe order necessarily produces SBs at momentum locations that are determined by, among others, the incommensurate ordering wave vectors. However, because of the short-range and fluctuating nature of the order, these SBs could be lack of intrinsic spectral weight and appear poorly defined in momentum space, and thus are hard to detect \cite{stripe:theory:SteveHowToDetect, stripe:theory:VojtaStripeReview}. A structural interpretation for the $(\pi,\pi)$ SB, which is associated with a commensurate wave vector and has an overall comparable spectral intensity with the MB, is hence not inconsistent with such picture.

\section{Summary}
We have presented the first comprehensive doping dependence study of the SB in La214 by ARPES. Its corresponding wave vector is close to $(\pi,\pi)$ and independent of doping. The SB at $x=0.07$ is shown to present a similar binding energy dependence with the MB in terms of the dispersion, quasi-particle momentum linewidth and spectral weight. Different from the Bi-based family, the intensity of SB at $E_F$ in La214 exhibits a systematic, monotonic change with doping, from being strong at low dopings to being negligible at high dopings. Essentially similar to the Bi-based case, we find an interpretation based on a unique form of structural distortion provides a better account for the origin of this SB than the magnetic alternative, although there seems to exist a semi-quantitative discrepancy from the experiment which warrants further theoretical examinations. Our results are inconsistent with a maximal SB intensity previously reported at $x=1/8$, and suggest that the manifestation of the stripe correlations in the band structure could be in a more subtle way.

\ack  R.-H.H. thanks the SGF for financial support. The work at Stanford and ALS is supported by the DOE Office of Basic Energy Sciences under contracts DE-AC02-76SF00515, DE-AC02-05CH11231 and a NSF grant DMR-0604701. The work at Tohoku is supported by Grant-In-Aid for Scientific Research (C) (20540342) and (B) (19340090) from the MEXT (Japan).

\appendix
\section{Doping dependent Fermi surface maps of La214}\label{app:A}

\begin{figure}[b]
\begin{center}
\includegraphics[angle=0, width=6 in]{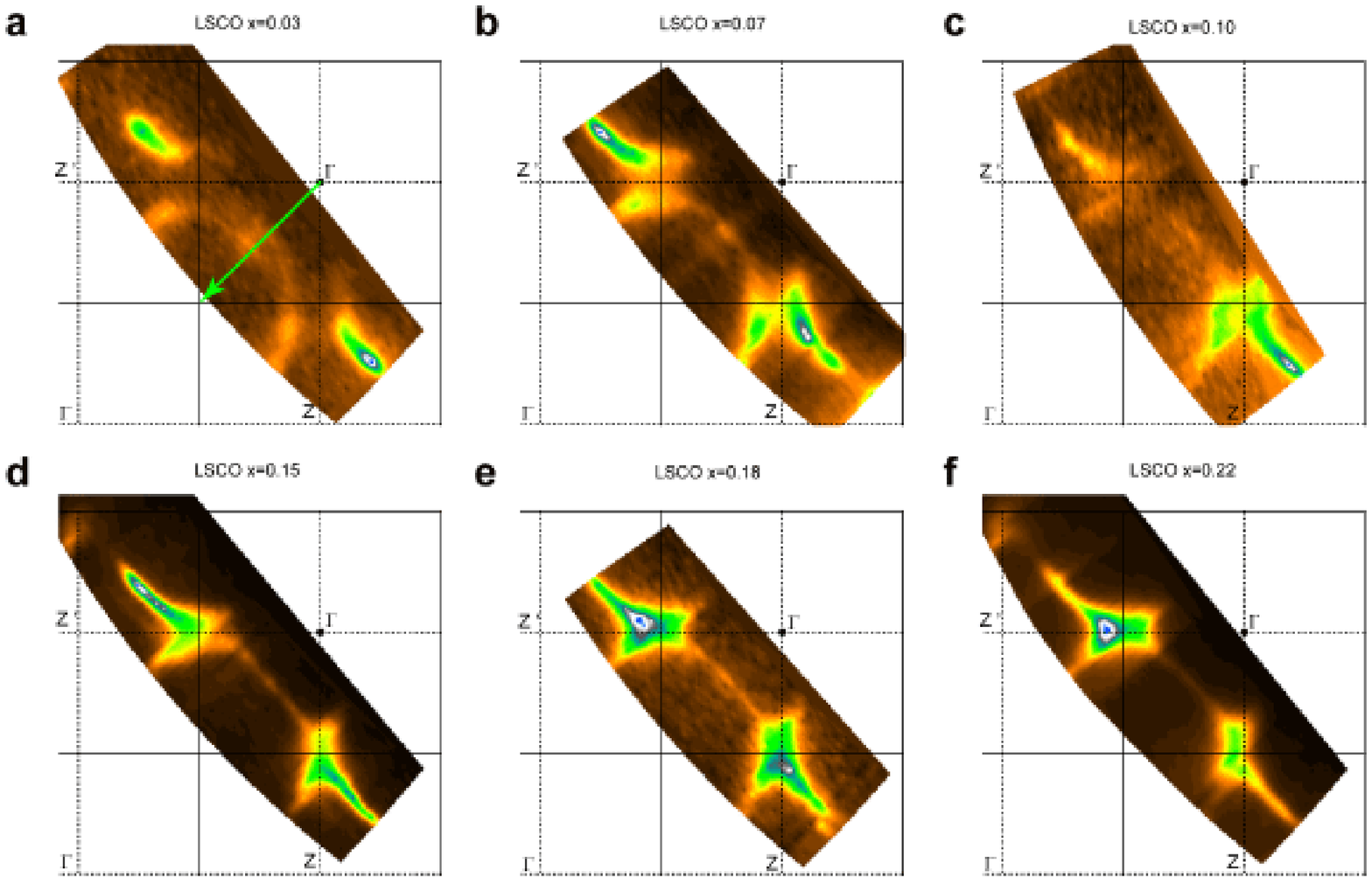}
\end{center}
\caption{FS maps of LSCO at different doping levels. Data were taken with SES-200 analyzer within 1999/11-2001/02. $h\nu$=55.5 eV. $E_F\pm$10 meV.}
\label{Fig. B200}
\end{figure}

\begin{figure}[b]
\hspace*{-1cm}
\includegraphics[angle=0, width=7 in]{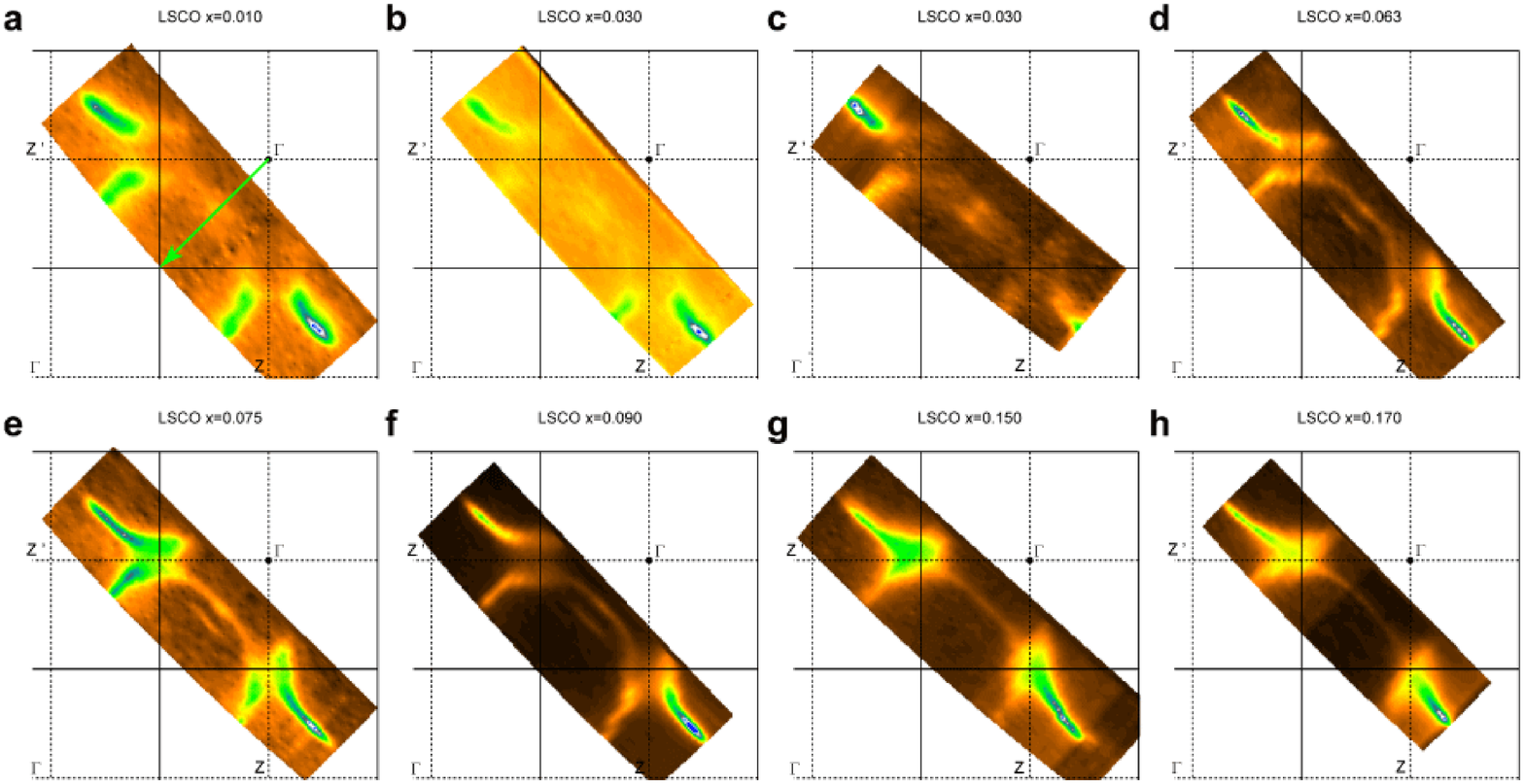}
\caption{FS maps of LSCO at different doping levels. Data were taken with SES-2002 analyzer within 2002/05-2004/04. $h\nu$=55 eV for \textbf{e}-\textbf{g} and $h\nu$=59.5 eV otherwise. $E_F\pm$10 meV.}
\label{Fig. B2002}
\end{figure}

\begin{figure}[t]
\begin{center}
\includegraphics[angle=0, width=6 in]{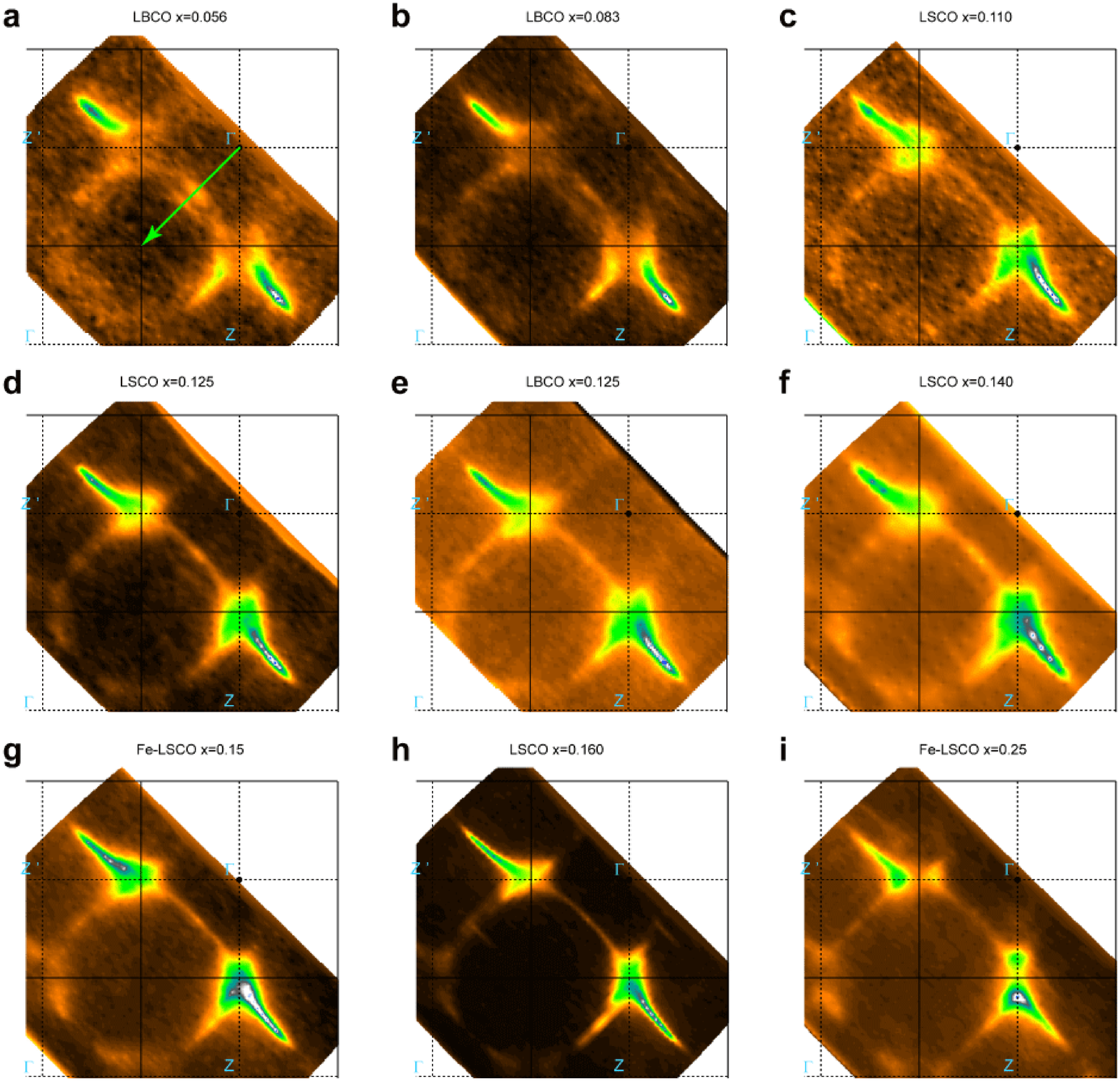}
\end{center}
\caption{FS maps of LBCO, LSCO and Fe-LSCO at different doping levels. Data were taken with SES-R4000 analyzer since 2006/04. $h\nu$=55 eV. $E_F\pm$10 meV. The data of LBCO-1/8 is reproduced from Ref. \cite{stripe:other:LBCO_twogap}, Fe-LSCO $x=0.25$ from Ref. \cite{stripe:other:FeLSCO25p} and the rest from Ref. \cite{stripe:other:WeaktoStrongCrossover}. Additional features, having an overall angular offset from the main one, in 1st BZs of Fe-LSCO $x=0.25$ are from another crystalline domain.}
\label{Fig. BR4k}
\end{figure}

\clearpage
\bigskip
\section*{References}
\bibliographystyle{unsrt}

\begin{thebibliography}{10}
\bibitem{Note_detwinned}
Note that our LSCO samples are not detwinned in terms of the orthorhombic distortions so a similar polarization dependence study as Ref. \cite{cuprates:Bi2212:Sbpipi_4} was not performed. In fact, according to that work, the current polarization condition for Cut 2 in Fig. \ref{Fig. 1}a should favor an anti-phase intensity distribution between the MB and SB which is not found in our experiments, presumably due to the different nature of distortions in La214 and/or the twinned nature of crystals.
\bibitem{Note_fit}
The FWHM asymmetry for the MB and SB as discussed in Section \ref{MDCanalysis} is not addressed here. Identical FWHM was assumed to reduce the number of fitting parameters. A two-Lorentzian fit with this constraint removed yielded similar results within error bars for cases where the SB intensity is strong, but it often gave unphysical results of the SB-MB midpoint for cases where noise dominates over the SB intensity.
\end{thebibliography}

\end{document}